\documentclass[aps,prl, superscriptaddress,showpacs,twocolumn,amssym]{revtex4}
\usepackage{latexsym}
\usepackage{graphicx}
\usepackage{rotating}
\usepackage{longtable}
\usepackage[usenames]{color}
\usepackage[normalem]{ulem}

\input colordvi

\begin{document}

\title{Stress induced stripe formation in Pd/W(110)}

\author{T.~O.~Mente\c{s}}
\affiliation{Sincrotrone Trieste S.C.p.A., Basovizza-Trieste 34012, Italy}

\author{A.~Locatelli}
\affiliation{Sincrotrone Trieste S.C.p.A., Basovizza-Trieste 34012, Italy}

\author{L.~Aballe}
\affiliation{CELLS-ALBA Synchrotron Light Facility, C3 Campus Universitat 
Aut\'{o}noma de Barcelona, 08193 Bellaterra, Barcelona, Spain}

\author{E.~Bauer}
\affiliation{Department of Physics, Arizona State University, Tempe, Arizona 85287-1504}

\date{\today}

\begin{abstract}
A stress-induced stripe phase of submonolayer Pd on W(110) is observed by
low-energy electron microscopy. The temperature dependence of the pattern
is explained by the change both in the boundary free energy and 
elastic relaxation energy due to the increasing boundary width.
The stripes are shown to
disorder when the correlation length of the condensed phase becomes
comparable to its period, while the condensate to lattice-gas
transition takes place at a higher temperature, as revealed by
low-energy electron diffraction.
\end{abstract}


\maketitle

It is well known that surface stress may give rise to periodic 
structural modulations~\cite{ibach97}. 
The most common examples are
surface recontructions on clean and adsorbate-covered crystal faces. 
In the presence of competing long-range
and short-range interactions, restructuring of
the surface can take place at mesoscopic length scales~\cite{alevanmea88}.

Several experiments have shown the formation of mesoscopic patterns as a result of
elastic interactions.
Examples include ($2\times1$)-O domains
on Cu(110)~\cite{kerniesch91}, ordered two-dimensional (2D) islands of 
Ag on Pt(111)~\cite{zepkrzrom95}, stripes of alternating dimer direction
on Boron doped Si(001)~\cite{jonpelhon96,hanbarswa97}, square patterns
of N/Cu(100)~\cite{ellreprou00}, stripe domains of Pb/Cu(111)~\cite{vanplabar03},
and Au stripes on W(110)~\cite{figleobar08}.
In all these cases, the length scales of the periodic modulations range
from a few to hundreds of nanometers, well above the atomic distances. 
This is understood within the analytical theory~\cite{alevanmea88},
which states that the period of a given stripe pattern 
depends exponentially on the ratio of the short-range (boundary) and
long-range (elastic) interaction energies. The period $D$ can be written as
\begin{equation}
\label{eq:eqn1}
 D\ =\ 2 \pi a\ e^{\frac{C_{1}}{C_{2}} + 1}\ \equiv \ w_0\ e^{\frac{C_{1}}{C_{2}} + 1},
\end{equation}
where $a$ is a microscopic length, $C_1$ is the formation energy of the stripe boundary,
and $C_2$ is the prefactor of the energy gained by 
the elastic relaxation due to the formation of stripes~\cite{alevanmea88}.
The prefactor of the exponent comes from a smearing of the stripe boundaries, 
and so is related to the boundary width.
This result is identical to that obtained for magnetic layers with dipolar
interactions~\cite{kaspok93,debmacwhi00}. We note that the correspondence between the dipolar
Ising lattice and the 2D lattice gas is one-to-one, as the elastic interaction energy
between lattice defects scales as $1/r^3$~\cite{peyvalmis99}.

Studies of mesoscopic patterns show that
the periodicity varies strongly with temperature. 
Moreover, there exists a transition temperature, at which
the stripe phase disappears. This temperature dependence has
been attributed to the reduction of the boundary free energy through
density fluctuations as the disordering temperature is approached~\cite{vanbarfei04}.
Earlier work on Ising spin lattices showed that the effect of fluctuations can be
introduced by an entropic term, which results in a temperature dependent
period for the magnetic stripes~\cite{czevil89}. 

\begin{figure*}[t]
\begin{center}
\includegraphics{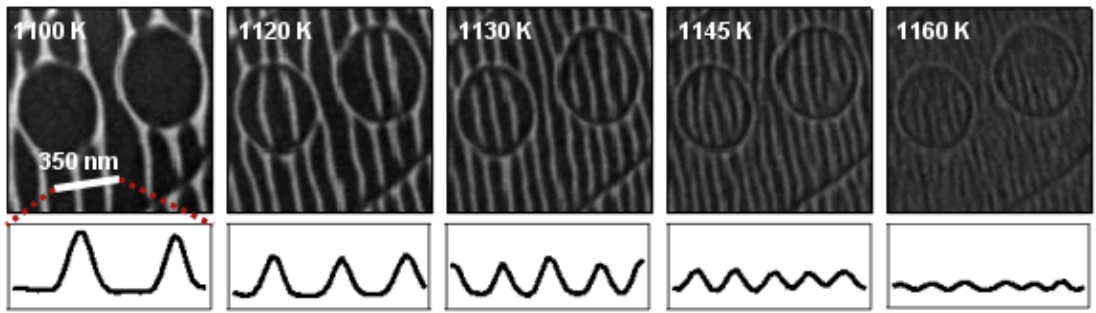}
\caption{(Color online) Monolayer Pd stripes on W(110) with increasing temperature. 
The stripe direction is along $[1\bar{1}0]$.
The condensed Pd islands appear bright at the electron energy used
(5.2~eV). The image size is 1~$\mu$m$^2$. 
A 350~nm profile across the stripes is shown below each image. Note that the
rectangular profile of the sharp boundaries is broadened due to the
resolution of the microscope. In all images, monoatomic substrate steps with circular shapes
can be recognized.}
\label{fig:Pd_stripes}
\end{center}
\end{figure*}

The density fluctuations, which reduce the boundary free energy, also increase
the boundary (or domain wall) width with increasing temperature~\cite{vinsarpor08}.
Gehring and Keskin took this broadening into consideration, and they argued
that the stripe disordering transition takes place when the boundary width becomes
half the stripe period~\cite{gehkes93}. 

All studies up to date
treat the temperature regime close to the transition separately, pointing to
a ``crossover'' as the boundary width becomes comparable to 
the stripe period (a discussion can be found in~\cite{kaspok93}).
In this paper, we report the observation of a stripe phase in sub-monolayer Pd films
grown on W(110). We demonstrate that the temperature dependence can be reproduced
throughout the whole temperature range by a scaling of the boundary energy and boundary width. 
Using low-energy electron microscopy (LEEM),
we show that the triangular step-flow patterns \cite{ababarloc04} observed below 1100 K
turn into stripes that run along the $[1\bar{1}0]$ direction above this temperature.
The period of the pattern sharply decreases from 200~nm to 57~nm 
with increasing temperature until the stripes disorder at around 1170 K.
The dependence of the period on
Pd coverage is much weaker contrary to the expectation from the
theory with sharp boundaries (a similar behaviour was found for
Au/W(110)~\cite{figleobar08}).

The thermodynamics of Pd/W(110) can be described as that of a lattice gas
with attractive interactions, 
and shows a condensate-gas transition with increasing temperature~\cite{kolbau85,kolbau96}.
Structurally, a Pd monolayer on W(110) shows one-dimensional pseudomorphism 
along the $[001]$ direction, whereas along $[1\bar{1}0]$ extra 
low-energy electron diffraction (LEED) spots are observable even
at the highest temperatures corresponding to the Pd lattice relaxed through periodic
dislocation lines~\cite{kolbau96}. This indicates a stress anisotropy with
a low surface stress along $[1\bar{1}0]$, readily explaining the
direction of the observed stripes in our study.

The LEEM measurements were performed with the SPELEEM
microscope at Elettra (Italy)~\cite{locabamen06}. 
The instrument allows real-time monitoring of the growth and 
evolution of adsorbate layers with a lateral resolution of 12~nm. 
The W(110) sample was cleaned by annealing at 1000 C in $2\times10^{-6}$~mbar oxygen,
followed by repeated high temperature flashes in ultrahigh vacuum to remove oxygen. 
The absence of oxygen was confirmed by the sharp, low background ($1\times1$) 
LEED pattern. Pd was deposited on the tungsten surface using
an e-beam evaporator at a rate of 0.2 monolayer per minute. 
The rate was calibrated via the time needed to complete the first monolayer at 800 C.

A series of LEEM images of alternating Pd stripes in condensed and lattice gas phases 
are shown in Fig.~\ref{fig:Pd_stripes}. The period of the pattern and the contrast
between the two phases sharply decrease with increasing temperature
up to about 1170 K, at which point the stripes fully disorder. 
The variation of the periodicity is facilitated by the high mobility
of the condensed regions. 
This leads to stripe fluctuations around the reversible stripe disordering
transition.

\begin{figure}
\begin{center}
\includegraphics[width=7.5cm]{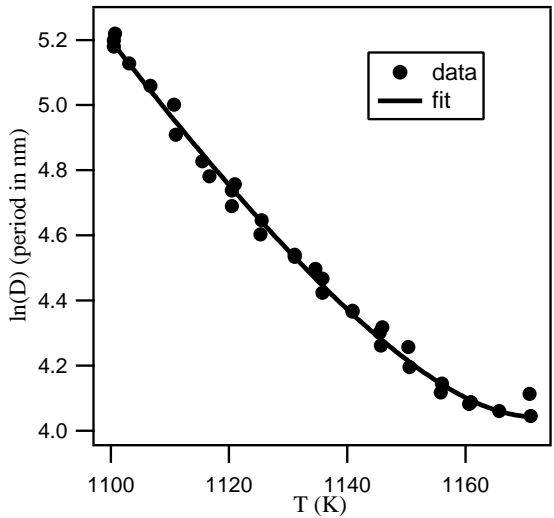}
\caption{Natural logarithm of the stripe period as a function of temperature.
The data points are shown as solid circles. The line corresponds to the fitting 
function in eq.~(\ref{eq:eqn5}).}
\label{fig:periodvsT}
\end{center}
\end{figure}

The natural logarithm of the stripe period, $D$, is displayed in Fig.~\ref{fig:periodvsT}
as a function of temperature. Except for temperatures very close to the stripe 
disordering transition, the logarithm of the period shows a linear trend, 
suggesting that the temperature dependence comes mainly from 
the energetic parameters in the exponent in  eq.~(\ref{eq:eqn1}). 
As we will discuss shortly, due to smearing of the stripe boundary,
the period levels off at a nonzero value close to the disordering transition.

As mentioned above, studies on the dipolar Ising lattice are
informative also for the behaviour of adsorbate stripes~\cite{note_Ising}.
Recent studies show that
very close to the disordering transition, the stripe profile can be approximated
by sinusoidals (keeping the lowest orders in a Fourier expansion)~\cite{porvatpes06}.
The result is a quadratic dependence of the period on temperature obtained by
minimizing the free energy:
\begin{equation}
\label{eq:eqn2}
 D = D_0\ \left[1 + c\ \left(1-\frac{T}{T_c}\right)^2\right],
\end{equation}
where $D_0$ is the period at the disordering temperature, $c$ is a constant given as a
combination of the interaction parameters, and $T_c$ is the disordering temperature.
Despite being a good approximation close to the disordering temperature, 
it breaks down away from the transition due to the implicit assumption that the width of
the stripe boundary is comparable to its period. 
Indeed, Fig.~\ref{fig:Pd_stripes} shows that at the
lower temperatures the stripe period is much larger than the boundary width,
and that the two phases (i.e. dark and bright stripes) have unequal widths.
Thus the lowest order components of a Fourier expansion are no longer
sufficient to describe the stripe profile.

Instead, the temperature dependence can be obtained from general scaling
laws governing the parameters in eq.~(\ref{eq:eqn1}). For the dipolar Ising lattice, 
the changes in the boundary free energy and boundary width have been identified  
as the determining factors~\cite{gehkes93}. The boundary free energy was found to decay
linearly with temperature due to the increase in entropy~\cite{fisfer67,vanbarfei04}.
The result is a scaling of the exponent in the stripe period:
\begin{equation}
\label{eq:eqn3}
 \frac{C^*_1}{C_2} = \frac{C_1}{C_2} \left(1 - \frac{T}{T^0_c}\right),
\end{equation}
where $C^*_1$ is the temperature dependent boundary free energy, and $T^0_c$ is the
transition temperature for the Ising model without the long-range dipolar or elastic
interactions. On the other hand, the boundary width scales as the correlation
length~\cite{gehkes93}. Using the critical exponent $\nu=1$ for the 2D Ising model,
\begin{equation}
\label{eq:eqn4}
 w = w_0 \left(1 - \frac{T}{T^0_c}\right)^{-1},
\end{equation}
where we define the unscaled boundary width as $w_0 \equiv 2 \pi a$. 
Combining eqs.~(\ref{eq:eqn1}), (\ref{eq:eqn3}) and (\ref{eq:eqn4}),
we can write the logarithm of the stripe period as
\begin{equation}
\label{eq:eqn5}
 ln(D) = \frac{C_1}{C_2} \left(1 - \frac{T}{T^0_c}\right) - ln\left(1 - \frac{T}{T^0_c}\right) 
         + ln(w_0) + 1 .
\end{equation}
The basic features in Fig.~\ref{fig:periodvsT} can now be understood.
The first term on the right hand side of eq.~(\ref{eq:eqn5}) dominates at the lower
temperatures giving the linear trend in $ln(D)$ vs $T$. The second term has
a mild temperature dependence away from the transition, but becomes 
increasingly important close to it. This corresponds to
a broadening of the stripe boundary, which slows down the variation in
the period and levels it off at the stripe disordering temperature.
 
The resulting fit to the experimentally determined stripe period
is shown as a solid line in Fig.~\ref{fig:periodvsT}.
The fitting parameters were found to be $w_0 = 1.95 \pm 0.04$~\AA, $T^0_c = 1201.7 \pm 2.6$ K and
the energy ratio $C_1 / C_2 = 39.8 \pm 0.8$. The scaling
gives excellent agreement with the data throughout the whole temperature range
for which stripes are observed. 
It should also be noted that near the disordering temperature, $T_c$, the quadratic
dependence in eq.~(\ref{eq:eqn2}) can be retrieved from the functional form
used for the fit (see Appendix).

The value of $T^0_c$ reveals an important effect. The scaling laws in
eqs.~(\ref{eq:eqn3}) and (\ref{eq:eqn4}) were found from the 2d Ising model.
They do not take into account the long-range
elastic interactions, which counteract the short-range interactions
between Pd atoms and favor disorder, lowering the transition
temperature from the pure Ising value, $T^0_c$, to the stripe
disordering temperature, $T_c=1170$~K.
Moreover, as predicted~\cite{stosin01}, we see that
the decrease of the transition temperature is given
by the ratio of the long-range and short-range energies:
\begin{equation}
\label{eq:eqn6}
 \left(1 - \frac{T_c}{T^0_c}\right) \ \approx \ \frac{C_2}{C_1}.
\end{equation}
In physical terms, the transition temperature is lower than the value expected
from a diverging correlation length, because the disorder sets in when the boundary
roughness (or the correlation length) becomes comparable to the stripe period. 
The same phenomenon is observed in finite size Ising lattices, in which the
disorder transition takes place when the fluctuations reach the system size. 
Such a reduction in the transition temperature was shown to scale inversely
with the finite size~\cite{fisfer67}. The period of Pd stripes at
the transition is approximately given by $D_0 = e^2 w_0 (C_1/C_2)$ from
Eqs.~(\ref{eq:eqn5}) and (\ref{eq:eqn6}), providing an analogous relation 
between the period and the lowering of the transition.

\begin{figure}
\begin{center}
\includegraphics[width=7.5cm]{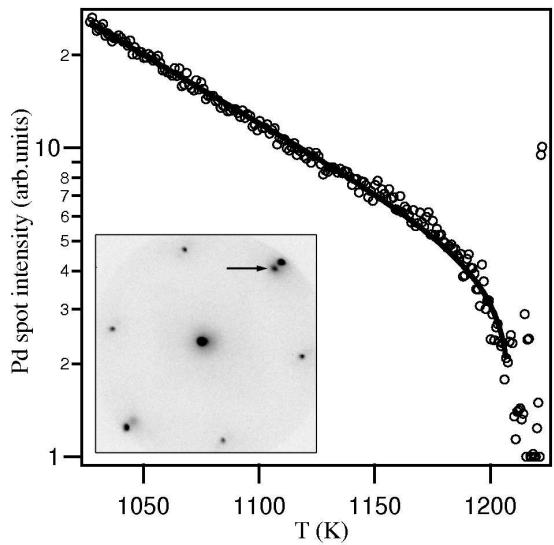}
\caption{The intensity of the Pd extra LEED spot (shown by the arrow in the inset) as
a function of temperature. Electron energy is 35 eV.
The solid line is a fitting function of the form 
$e^{E_k/kT} (1-T/T^0_c)^{2\beta}$, where $\beta=1/8$ is the critical exponent for the order
parameter in the 2D Ising model. $E_k$ is the energy of sublimation from the condensate 
to the lattice gas phase.}
\label{fig:LEED}
\end{center}
\end{figure}

The actual condensate to lattice gas phase transition, which should take place
at $T^0_c$, can be observed in the intensity of the Pd diffraction spots shown
in the inset of Fig.~\ref{fig:LEED}. In agreement with the discussion above,
we observe that the crystalline order of the Pd layer persists 
after the stripe contrast vanishes at $T_c$. 
A fit to the temperature dependence of the diffraction intensity
(see Fig.~\ref{fig:LEED}) 
gives a condensate to lattice gas transition at $T^0_c=1210 \pm 5$~K, only
slightly higher than the value extracted from Eq.~(\ref{eq:eqn5}). 
The intermediate phase, between $T_c$ and $T^0_c$, 
does not have the long-range stripe order, however it consists of
condensed islands buried in the lattice gas phase as discussed in~\cite{boomacwhi95}.

The exponential decay of the Pd diffraction spot intensity with temperature
up to about 1170 K is much faster than that expected for a physically realistic
Debye-Waller factor. It is rather due to the decrease of the number of atoms
in the crystalline phase by sublimation into the lattice gas. Consequently,
the energy in the exponential, $E_k=0.88$~eV, corresponds to the removal of
Pd atoms from the kink sites of the stripe boundaries. 
This translates into a boundary energy of $C_1 = 197$~meV/\AA\ along $[1\bar{1}0]$.
The value of the boundary energy, along with $C_1/C_2 = 39.8$,
results in an elastic energy parameter $C_2 \approx 5.0$~meV/\AA.
Using the bulk elastic constants for tungsten, we obtain the surface stress
change across the Pd step to be $\Delta\tau^{[001]} \approx 4.6$~N/m
(assuming an isotropic elastic response for
the tungsten substrate according to~\cite{leobarkel05}).
This is in very good agreement with density-functional theory calculations,
which suggest that the tungsten surface stress along $[001]$, 5.26~N/m~\cite{menstobin08},
is reduced nearly to zero upon Pd adsorption~\cite{private_stojic}.

Finally, we note that no orientational melting of the stripes is observed.
With increasing temperature, contrast between the condensate and gas stripes
decreases monotonically, until it disappears at $T_c$. However, the direction
of the pattern is preserved all the way up to $T_c$ as shown in Fig.~\ref{fig:Pd_stripes}.

In conclusion, we have shown that stress-induced stripes of Pd on W(110)
behave according to the scaling laws for the 2D Ising model with increasing
temperature. Using low-energy electron diffraction and microscopy, we have
demonstrated that the condensate to lattice gas transition takes place
at a slightly higher temperature than the disordering of the stripes. In analogy
with the lowering of the transition temperature in the finite-size Ising lattice,
this difference in the disordering temperature of the atomic and mesoscopic scales
is explained by the loss of stripe order when the correlation length becomes
comparable to the stripe period. 

\acknowledgements
We thank Nata\v{s}a Stoji\'{c} for making available the results of her
surface stress calculation on Pd/W(110) before publication.

\renewcommand{\theequation}{A.\arabic{equation}}
\setcounter{equation}{0}
\paragraph*{Appendix.}
Eq.~(\ref{eq:eqn2}) can be obtained by expanding eq.~(\ref{eq:eqn5}) near
the stripe disordering temperature, $T_c$. We first rewrite the scaling expression
for a temperature slightly below $T_c$:
\begin{equation}
\label{eq:eqnAppendix1}
  D(T_c - \delta T) = \frac{ew_0}{1-\frac{T_c-\delta T}{T^0_c}} \ 
                     e^{\frac{C_1}{C_2}\left(1-\frac{T_c-\delta T}{T^0_c}\right)}.
\end{equation}
Using eq.~(\ref{eq:eqn6}) we can expand both the denominator and the exponent
in powers of $(C_1/C_2)(\delta T/T^0_c)$. The result has the same form
as in eq.~(\ref{eq:eqn2}):
\begin{equation}
\label{eq:eqnAppendix2}
  D(T) \approx D_0 \left[1 + \frac{1}{2} \left(\frac{C_1}{C_2}\right)^2 
\left(1-\frac{T}{T_c}\right)^2 \right] + O(..)^3,
\end{equation}
where $D_0$ is the period at $T_c$, and $T = T_c - \delta T$.


\end{document}